\documentstyle[referee]{mn}
\newcommand{\reference}{\bibitem}
\def\beq{\begin{equation}}
\def\eeq{\end{equation}}
\def\bey{\begin{eqnarray}}
\def\eey{\end{eqnarray}}
\def\beqarray{\begin{eqnarray}}
\def\eeqarray{\end{eqnarray}}

\def\kpc{\,{\rm {kpc}}}

\def\kms{\,{\rm {km\, s^{-1}}}}

\def\v200{V_{200}}

\def\Log{\,{\rm Log}\,}

\input epsf
\begin{document}

\title[]
{The Fundamental Plane of Spiral Galaxies: Theoretical
Expectations}
\author[]
{Shiyin Shen$^{1,2,3}$, H. J.
Mo$^2$, Chenggang Shu$^{1,2,3}$
\thanks {E-mail: ssy@center.shao.ac.cn,
hom@mpa-garching.mpg.de, cgshu@center.shao.ac.cn} \\
$^1$ Shanghai Astronomical Observatory, Chinese
Academy
of
Sciences, Shanghai 200030, China\\
$^2$ Max-Planck-Institut
f$\ddot{u}$r Astrophysik,  Karl-Schwarzschild-Strasse
1,
85748
Garching, Germany\\
$3$ Joint Lab of Optical
Astronomy, Chinese Academy of Sciences, China\\}
\date{Accepted ........      Received .......;
in original form .......}

\pubyear{2001}
\maketitle
\begin{abstract}
Current theory of disk galaxy formation is used  
to study fundamental-plane (FP) type of relations 
for disk galaxies. We examine how the changes 
in model parameters affect these relations and 
explore the possibility of using such relations to
constrain theoretical models. The distribution 
of galaxy disks in the space of their fundamental
properties are predicted to be concentrated in a plane, with the
Tully-Fisher (TF) relation
 (a relation between luminosity $L$ and maximum rotation 
velocity $V_m$) being an almost edge-on view. Using 
rotation velocities at larger radii generally leads to
larger TF scatter. In searching 
for a third parameter, we find that both the
disk scale-length $R_d$ (or surface brightness) 
and the rotation-curve shape are correlated 
with the TF scatter. The FP relation in the 
$(\Log L, \Log V_m, \Log R_d)$-space
obtained from the
theory is $L\propto R_d^{\alpha'} V_m^{\beta'}$, 
with ${\alpha'}\sim 0.50$ and ${\beta'}\sim 2.60$, 
consistent with the preliminary result we obtain 
from observational data. Among the model parameters we
probe, variation in any of them can generate significant
scatter in the TF relation, but the effects of the 
spin parameter and halo concentration can be reduced 
significantly by introducing $R_d$ while the scatter
caused by varying $m_d$ (the ratio
between disk mass and halo mass) is most effectively
reduced by introducing the parameters which describes the
rotation-curve shape.
The TF and FP relations combined should therefore 
provide useful constraints on models of galaxy
formation. 
\end{abstract}


\section{Introduction}

    Spiral galaxies are characterized by their
flattened disks
(with approximately exponential profiles) and nearly
flat rotation curves. The main observational parameters
that describe the overall properties of a
spiral galaxy are the central surface brightness
$\mu_0$, the disk scale-length $R_d$ and the characteristic
rotation velocity (e.g. the maximum rotation velocity $V_m$). The
observed disk population
covers a large range in these parameters, with $\mu_0$
varying from $\sim 24$ (low surface brightness galaxies) to
$\sim 20\,\rm{mag\,arcsec^{-2}}$, $R_d$ from $\sim 0.1$ to $\sim
10\kpc$, and $V_m$ from $\sim 50$ to $\sim 300\kms$. Despite
the diversities, spiral galaxies obey a well-defined scaling relation 
between their total luminosity $L$ and maximum
rotation velocity $V_m$.
This relation (called the Tully-Fisher relation,
hereafter TF relation) is usually expressed in the form 
\begin{equation}
L=A V_m^{\beta'}\,,
\end{equation}
where $\beta'$ is the slope, and $A$ is the zero-point.
The observed TF relation is quite tight.
For example, in the $I$-band, the scatter in luminosities
for a fixed $V_m$ is only about 0.38 magnitude 
(Giovanelli et al. 1997). Thus, the TF relation 
can be used as a relative distance indicator for
spiral galaxies. It is still unclear whether the scatter in
the observed TF relation is dominated by observational
errors or by intrinsic variations. Since intrinsic
scatter must exist to some degree, a natural question
is whether there is a third parameter that correlates
with the residual scatter of the TF relation. The situation is
quite similar to that for elliptical galaxies. 
Although elliptical galaxies are found to obey the
Faber-Jackson relation (a relation between luminosity $L$ and 
central velocity dispersion $\sigma$), the
fundamental-plane (hereafter FP) relation
(with the introduction of a third parameter, the
effective radius $r_e$) has much smaller scatter 
(Djorgovski \& Davis 1987; Dressler et al. 1987). 
For disk galaxies, due to the broad
distribution of scale-length $R_d$, one obvious
choice of the third parameter is $R_d$ or the central
surface brightness $\mu_0$ ($\mu_0$ and
$R_d$ are related by $L=2\pi\mu_0 R_d^2$ for an
exponential disk). 
On the other hand, since disk galaxies also present
diversities in their 
rotation-curve shapes (Persic, Salucci \& Stel 1996),
 another possible choice may be 
a parameter that characterizes the rotation-curve
profile.

 Much effort has been made in searching for this third
parameter
(e.g. Kodaira 1989; Han 1991; Tully \& Verheijen 1997;
Willick et al. 1997; Courteau \& Rix 1999; Willick 1999), 
but the results are still inconclusive.
While most of the studies cited above concluded that
the third parameter is not crucial, some of them did
find evidence for the existence of a third parameter  
(e.g. Kodaira 1989; Willick 1999). 
More recently, Koda, Sofue \& Wada (2000a) obtained a
scaling plane in the form of $L\propto (VR)^{\beta'}$ from both 
observational data and numerical simulations. 
But the radius $R$ in this relation is not really a
new degree of freedom, because it is bounded to the
rotation velocity $V$ with the same power index $\beta'$. 

 From theoretical side, there are many attempts to
understand the formation of disk galaxies in the framework provided
by current cosmogonic models (e.g. CDM models: Fall \& Efstathiou
1980; White \& Frenk  1991; Dalcanton, Spergel \&
Summes 1997; Mo, Mao \& White 1998, hereafter MMW).
The current standard scenario of disk formation
assumes that galaxy disks form as a result of gas cooling in dark
matter haloes. Detailed modeling shows that such models are
quite successful in interpreting observational data, 
especially the observed TF relation
(MMW; van den Bosch 1998, 2000; Avila-Reese, Firmani
\& Hernandez 1998; Weil, Eke \& Efstathiou 1998; 
Heavens \& Jimenez 1999; Mo \& Mao 2000;  
Navarro \& Steinmetz 2000; Koda, Sofue \& Wada 2000b; 
Zou \& Han 2000; Cole et al. 2000; 
Buchalter, Jimenez \& Kamionkowski 2001).
In the CDM cosmogonies, the characteristic 
rotation velocity of a disk galaxy is determined
largely by the potential of the dark matter while the
luminosity of the disk is given by the amount of stars (baryons) 
in the disk. The observed TF relation therefore
implies a tight relation between the disk mass and the depth
of the dark halo potential well, and its scatter may be
used to constrain models of galaxy formation.
  
Our main goal in this paper is  to examine whether 
FP-like relations exist for the disk population based on current
theoretical models of disk formation, and to provide some
theoretical guidelines for the search of FP-like relations.
Moreover, we explore the
possibility of using such relations as a tool to
understand galaxy formation by examining the response of the theoretical
scaling relations to the changes of model parameters.

The structure of the paper is as follows.
Section 2 presents the model and summarizes the 
model parameters. In Section 3, we search for the 
theoretical FP relation and examine its responses
to the variations of model parameters.
In Section 4, some preliminary results obtained from 
observational data are presented and compared with 
our model predictions. Finally, in Section 5, we make 
further discussions and summarize our main
conclusions.

\section{Theoretical model}
\label{theory}

\subsection{Disk Formation}

Our model here follows that presented in Mo, Mao \&
White (1998, hereafter MMW). We refer the reader to that
paper for details; 
here we only introduce the main ingredients relevant
to our analyses.

The initial density profiles of dark haloes are
assumed to take the universal form, 
\begin{equation}
 \rho(r)=\frac{\rho_{crit} \delta_0}
  {(r/r_s)(1+r/r_s)^2}\,,
\end{equation}
where $r_s$ is a characteristic radius, $\rho_{crit}$ is the critical 
density and $\delta_0$ is a constant (Navarro, Frenk \& White 1997, 
hereafter NFW). We define the radius of a dark halo to be
$r_{200}$ within which the mean density is 200 times the
critical density, and represent the characteristic radius 
by the concentration, $c\equiv r_{200}/r_s$. It is then 
easy to show that 
\begin{equation}
\delta_0={200\over 3}{c^3\over\ln(1+c)-c/(1+c)}\,.
\end{equation}
The radius $r_{200}$ and mass $M$ of a halo can be 
expressed in terms of its circular velocity $V_c$ as  
\begin{equation}
  r_{200}=\frac{V_c}{10H(z)}\,,~
M=\frac{{V_c}^2r_{200}}{G}\,,
\end{equation}
where $G$ is the gravitational constant, and $H(z)$ is
the Hubble constant at redshift $z$.
 We relate the initial angular momentum $J$ of a halo
to its spin parameter $\lambda$ through the definition
\begin{equation}
\lambda=J|E|^{1/2}G^{-1}M^{-5/2},
\end{equation}
where $E$ is the total energy of the halo. A dark halo
is therefore described by three parameters: the
circular velocity
$V_c$, the concentration $c$ and the spin parameter
$\lambda$.
As a result of dissipative and radiative processes,
the gas 
component gradually settles into a disk. We assume the
disk mass to be a fraction $m_d$ of the halo mass, and
the disk angular momentum to be $j_d$ times  $J$.
Thus, two other parameters, $j_d$ and $m_d$ are 
introduced to
relate the halo properties to the properties of the
final disk. Following MMW, we also assume
that disks have exponential surface density profiles
with a constant mass-to-light ratio $\Upsilon$, 
and that dark haloes respond to disk growth
adiabatically. 
Under these assumptions, we can obtain the luminosity 
$L$, the disk scale length $R_d$ and the rotation
curve $V(R)$ for
a given set of model parameters. Specifically,
\begin{equation}
L=\frac{m_dM}{\Upsilon}\,,
~ 
R_d=\frac{1}{\sqrt
2}\left(\frac{j_d}{m_d}\right)\lambda{r_{200}}f_r\,,
~V(R)=V_cf_v(R)\,,
\end{equation}
where $f_r$ and $f_v$ are factors which depend on halo
profile and disk action. 
We use the same procedure outlined in MMW to calculate
$f_r$ and $f_v$
as functions of $c$, $j_d$, $m_d$ and $\lambda$.

\subsection{Model Parameters}

 As discussed above, five parameters 
($V_c$, $\lambda$, $c$, $m_d$ and $j_d$) need to be 
specified for individual haloes in order to predict
the properties of the disks that form within them. In this
subsection we describe one by one how these parameters are 
chosen in our modelling.

\subsubsection{Halo circular velocity $V_c$}

The distribution of halo masses is derived from the Press-Schechter
formalism (Press \& Schechter 1974, hereafter PS) applied to  
the $\Lambda$CDM cosmogony (with mass density parameter
$\Omega_0=0.3$, cosmological constant $\Omega_{\Lambda} =0.7$,
Hubble's constant $h=0.7$, perturbation-spectrum normalization
$\sigma_8=1$). We select the host haloes of disks with circular
velocities in the range from $50\kms$ to 300$\kms$ at redshift
$z=0$. The choice of cosmogony does not change our main results,
because we are not interested in the exact distribution of 
galaxies with respect to mass. 

\subsubsection{Halo spin parameter $\lambda$}

N-body simulations show that the distribution of halo
spin parameter $\lambda$ can be approximated by a
log-normal function,
 \begin{equation}\label{p_of_lambda}
p(\lambda)\,d\lambda=\frac{1}{\sqrt{2\pi}\sigma_{\ln\lambda}}
  \exp\left[-\frac{\ln^2(\lambda/\bar{\lambda})}{2\sigma_{\ln\lambda}^2}\right]
  \frac{d\lambda}{\lambda},
\end{equation}
with $\bar{\lambda}$=0.05 and
$\sigma_{\ln\lambda}=0.5$. 
This distribution is found to be quite independent of
cosmology and of halo mass (Warren et al. 1992; Lemson \&
Kauffmann 1999). Syer, Mao \& Mo (1999) obtained a similar
distribution function for $\lambda$
from the observational data of disk sizes. In this
paper, we use equation (\ref{p_of_lambda}) for the distribution of $\lambda$.

\subsubsection{Halo concentration $c$}

The halo concentration, $c$, may depend on halo
formation history and cosmogony. Simulations show that the 
circular-velocity dependence of the NFW concentration
$c$ can be approximated by
\begin{equation}\label{c_NFW}
c_{\rm NFW}\approx c_{\rm NFW *} (V_c/100\kms)^{-1/3}
\end{equation}
(see NFW). 
Using $N$-body simulations, Jing (2000) found that the
distribution 
of $c$ for the majority of his simulated haloes
can be described by a log-normal function with a mean
$\bar{c}$ slightly smaller than that of the NFW
result:
\begin{equation}\label{p_of_c} 
p(c')dc'=\frac{1}{\sqrt{2\pi}\sigma_{\ln c'}}
\left[-\frac{\ln^2(c'/\bar{c}')}
{2\sigma_{\ln c'}^2}\right]d\ln c'\,,
~c'\equiv c/c_{\rm NFW}\,,
\end{equation}
where $\bar{c}'=0.85$ and $\sigma_{\ln c'}=0.25$. 
In this paper, we take $c_{\rm NFW *}=10$
as is given in the original NFW paper, use equation
(\ref{c_NFW}) to obtain $c_{\rm NFW}$ and equation (\ref{p_of_c}) for the
distribution of $c$.

\subsubsection{The mass ratio $m_d$}

This parameter represents the ratio between the disk
mass and the total halo mass. The processes by which gas cools
into a disk 
to form stars are complicated and not well understood 
at the present time. Nevertheless, some limits
can be set. First, the value of $m_d$ should be
smaller than
the overall baryon fraction in the universe
$\Omega_B/\Omega_0$,
which is about 0.1 for a low-density universe with
$\Omega_0=0.3$
and $h=0.7$,  and about 0.05 for an Einstein-de Sitter
universe with $h=0.5$, according to the cosmic nucleosynthesis
theory. Second, the value of $m_d$ should not be much smaller
than 0.01, because 
disks with lower gas content may not be able to form
sufficient amount of stars to be included in a TF sample.
Thus, we set the range of $m_d$ to
be from 0.01 to 0.1 and take 0.05 as the typical
value. We will also consider a case where $m_d$ changes 
systematically with $V_c$. 

\subsubsection{The angular-momentum ratio $j_d$}

This parameter characterizes the ratio between the
specific angular
momentum of the disk and that of the dark halo. 
Although disk material may have the same initial
specific angular momentum as the dark matter, little is known
about the transfer of angular momentum between gas and
dark matter in subsequent evolutions. From earlier
theoretical considerations (e.g. MMW) we know that $j_d\approx
m_d$ is required in order to ensure the predicted disk
sizes to match observations. In our model, we allow $j_d$ 
to have $50\%$ chance to be in the range
$0.5m_d$--$1m_d$,
and another $50\%$ chance to be in the range  
$1m_d$--$2m_d$. Thus, we adopt for $j_d$ a simple
distribution function of the form:
\begin{equation}\label{p_of_jd} 
p(j_d)d j_d=\left\{\begin{array}{ll}  
   d j_d/m_d &\mbox {for $0.5\,m_d\le j_d<m_d$}\\
   d j_d/(2m_d) &\mbox {for $m_d\le j_d<2\,m_d$}
\end{array}\right. \,.
\end{equation}

\subsection{Monte-Carlo Realization
\label{Monte_Carlo}}

In order to search for a FP relation for theoretical
disks, we need to choose three observable parameters to define
the plane. We take the luminosity $L$ and a characteristic 
rotation velocity $V$ as the two fundamental quantities. 
In most of our discussions, we use the maximum velocity $V_m$ 
as the characteristic velocity, but we also discuss the effects 
of using rotation velocities at other radii. 
For the third quantity, we try two different choices 
based on the discussion in the previous section: the
disk scale-length $R_d$ and a rotation-curve shape
parameter $\Gamma$ (to be defined below). 

The procedure for obtaining the FP is as follows. We
first generate a Monte-Carlo sample of 1000 haloes 
from the PS formalism, with $V_c$ in the range between 50 
and $300\kms$. We then assign to each halo the other four 
parameters by Monte-Carlo simulations. Finally we use the model 
of MMW to predict the properties of each model galaxy. 
As in MMW, we exclude unstable disks with
\begin{equation}\label{epsilon_m}
\epsilon_m\equiv\frac{V_m}{{(GM_d/R_d)}^{1/2}}
< 0.9\,,
\end{equation}
where $M_d=m_d M$ is the gas mass of the disk 
(e.g. Christodoulou, Shlosman \& Tohline 1995). 
On the other hand, galaxies with very low density may not form stars.
We therefore also exclude disks with Toomre parameter
\begin{equation}\label{Toomre_Q}
Q\equiv \frac{\sigma\kappa}{\pi G\Sigma}>1.4\,,
\end{equation}
where $\sigma$ is the gas velocity dispersion, $\kappa$ is the 
epicyclic frequency, and $\Sigma$ is the mass surface  density 
(Kennicutt 1989).  
In our model, we use the average surface mass density 
inside the half-mass radius (about $1.68 R_d$) to 
replace $\Sigma$, and adopt $\sigma=6\kms$.
We assume that all disks have the same mass-to-light ratio 
and take $\Upsilon\approx1.7h$ in the $I$-band (Bottema 1997).
All the results are for disks at $z=0$ and 
we take $h=0.7$ whenever an explicit value of $h$ is used.

\section{Model Predictions}

\subsection{Results Based on Maximum Rotation
Velocity and Disk Scale-length}

In this section, we try to find the FP of disk galaxies
in the $(\Log L$-$\Log V_m$-$\Log R_d)$-space. The reason is
clear: while $L$
and $V_m$ are the two parameters in the conventional TF
relation, the choice
of $R_d$ as the third parameter is motivated by the
fact that disk
galaxies cover a large range of $R_d$. We define the FP as
\begin{equation}\label{FPfit}
M_I=\alpha\,\Log R_d({\kpc})+\beta\,\Log V_m
({\kms})+\gamma,
\end{equation}
where $M_I$ is the absolute magnitude of the disk in
the $I$-band, $\alpha$, $\beta$ and $\gamma$ are constants which are
obtained by a least-square fit of equation (\ref{FPfit}) to the Monte-Carlo
data. In the disk model outlined above, there are four parameters,
$\lambda$, $m_d$, $j_d$ and $c$, whose effects on the FP relation
need to be evaluated. If all these four parameters are set to be
constant, the predicted TF relation is a line with no scatter.
The scatter in any one of these four parameters can cause
scatter in the TF relation (see e.g. MMW, Mo \& Mao
2000). The first question is whether some of the scatter can be
eliminated or reduced in the FP relation. 
In order to show this, we first construct four 
samples where scatter is allowed only in one of the
four model parameters while the other parameters taking
their typical values (see Table 1).  
In these cases, we ignore the selection criteria 
given by equations (\ref{epsilon_m}) and (\ref{Toomre_Q}),
and so some of the model galaxies in these samples 
may not correspond to any realistic disks.
The scaling relations are shown in Figure 1, while 
the  fitting results  of these relations are listed in Table 1. 
The scatters on both the TF relation and the FP relation are
represented by the root-mean-square deviation ($rms$) of $M_I$.

\begin{figure}
\epsfysize=14.0cm 
\centerline{\epsfbox{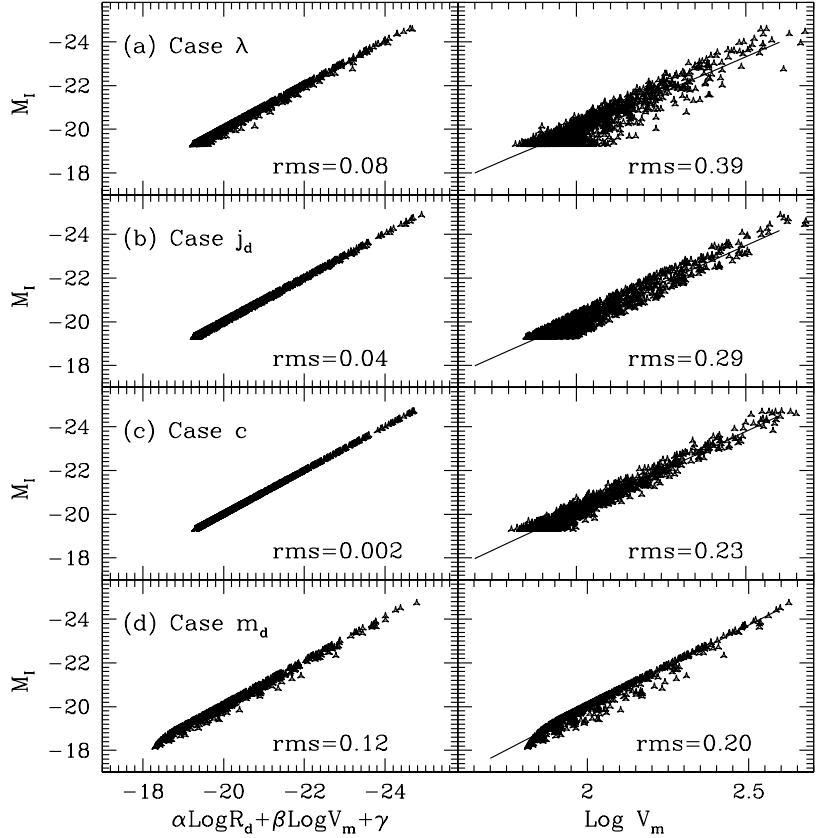}} 
\caption{The effects of changing individual model parameters
(as indicated in the panels) on the TF and FP relations.
(a), (b), (c) and (d) show the effects of $\lambda$, $j_d$, $c$ and
$m_d$ respectively. In each case, the left panel
shows the FP relation (the magnitude $M_I$ as a function of
$\alpha \Log R_d+\beta\Log V_m+\gamma$, where the values 
of $\alpha$, $\beta$ and $\gamma$ are listed in Table 1), whereas 
the right panel shows the corresponding TF relation
(the solid line being the linear regression). The
scatter in $M_I$ is represented by the $rms$ deviation.} 
\label{fig1}
\end{figure}

Comparing the scatter of the TF relation and the 
corresponding FP relation, we see that the introduction
of $R_d$ as the third parameter is very effective in reducing the 
scatter caused by $\lambda$, $j_d$ and $c$ but less 
so in reducing the scatter caused by $m_d$.
Notice that the FPs in different cases are tilted 
significantly with each other, except cases of
varying $\lambda$ and $j_d$ (see Table 1).   

\begin{table}
\caption{The effects of changing individual model parameters
on the TF relation and the FP relation defined by equation 
(\ref{FPfit}). Column 1 lists the name of each case, 
columns 2 to 6 give the model parameters (which are either a single number, 
or a range, or an equation of distribution), 
columns 7 to 9 show the fitting results for the FP
relation, while columns 10 and 11 list the scatter of 
the magnitude $M_I$ in the FP and TF relations, respectively}
\begin{tabular}{lcccccrrrrr} \hline
Case & $\lambda$ & $j_d$ &   $c$ & $m_d$& $\Upsilon$ &
$\alpha$ & $\beta$
& $\gamma$ & $\sigma_{\rm FP}$  &  $\sigma_{\rm TF}$
\\
\cline{1-11}
 $\lambda$ & eq.\,(7) & 0.05 & 8.5 &  0.05   &  $1.7h$
& $-1.29$ & $-6.16$ & $-7.21$ & 0.08 & 0.39 \\
 $j_d$ & 0.05 & eq.\,(10) &   0.05 & 8.5 &    $1.7h$
&$-1.21$ & $-6.27$ & $-7.05$  & 0.04 & 0.29 \\
 $c$ & 0.05 & 0.05 & eq.\,(9) & 0.05 & $1.7h$
& $-2.89$ & $-4.61$ & $-9.83$  & 0.002 & 0.23 \\
 $m_d$ & 0.05 & $m_d$ & 8.5 & 0.01--0.1 & $1.7h$
&1.11 & $-8.53$ & $-3.23$  & 0.12 & 0.20 \\
\hline
\end{tabular}
\end{table}

After an preview of the effects of individual model
parameters on the scaling relations, we now examine
cases where scatter is allowed in more than one model
parameters. For clarity, we list all cases in Table 2.
In all these cases, the selection criteria 
given by equations (\ref{epsilon_m}) and (\ref{Toomre_Q})
are imposed.

\begin{table}
\caption{The predicted TF ($M_I$-$\Log V_m$) 
and FP ($M_I$-$\Log V_m$-$\Log R_d$) relations in
various cases (Case VI will be discussed in Subsection 3.4).
 The table structure is the same as Table 1.
}
\begin{tabular}{ccccccrrrrr} \hline
Case & $\lambda$ & $j_d$ & $c$ & $m_d$ &  $\Upsilon$ &
$\alpha$ & $\beta$
& $\gamma$ & $\sigma_{\rm FP}$  &  $\sigma_{\rm TF}$
\\
\cline{1-11}
I & eq.\,(7) & eq.\,(10) &   8.5 & 0.05   & $1.7h$
&$-0.95$ & $-6.54$ & $-6.64$ & 0.03 & 0.21 \\
II & eq.\,(7) & eq.\,(10) &   eq.\,(9)& 0.05    & $1.7h$
&$-1.22$ & $-6.43$ & $-6.65$  & 0.15 & 0.35 \\
III & eq.\,(7) & eq.\,(10) & 8.5 &0.01--0.1 &  $1.7h$
& $-1.27$ & $-6.35$ & $-6.79$  & 0.32 & 0.44 \\
IV & eq.\,(7) & eq.\,(10) & eq.\,(9) &0.01--0.1 &  $1.7h$
&$-1.48$ & $-6.35$ & $-6.61$  & 0.34 & 0.50 \\
V & eq.\,(7) & eq.\,(10) &eq.\,(9) & 0.01--0.1 & eq.\,(15)
& $-0.79$ & $-6.47$ & $-6.85$  & 0.33 & 0.38\\
VI & eq.\,(7) & eq.\,(10) & eq.\,(9) & eq.\,(18) &  $1.7h$
&$-1.09$ & $-8.11$ & $-2.90$  & 0.22 & 0.35 \\
\hline
\end{tabular}
\end{table}

\subsubsection {Effects of $\lambda$ and $j_d$}

 Since the TF relation is, for a constant disk
mass-to-light
ratio, a relation between the disk mass and the
maximum rotation
velocity, difference in the disk angular momenta
may cause scatter in the TF relation because, for a
given disk mass, the contribution by the disk to the
rotation velocity depends on disk angular momentum. The disk
specific angular momentum is characterized by the
product $\lambda j_d$ and so we examine the effects of these
two parameters together. We generate a Monte-Carlo
sample, with $\lambda$ and $j_d$ following the
distributions described in equations (\ref{p_of_lambda})
and (\ref{p_of_jd}),
while keeping $m_d=0.05$ and $c=8.5$ (see Case I in
Table 2).

\begin{figure}
\epsfysize=14.0cm 
\centerline{\epsfbox{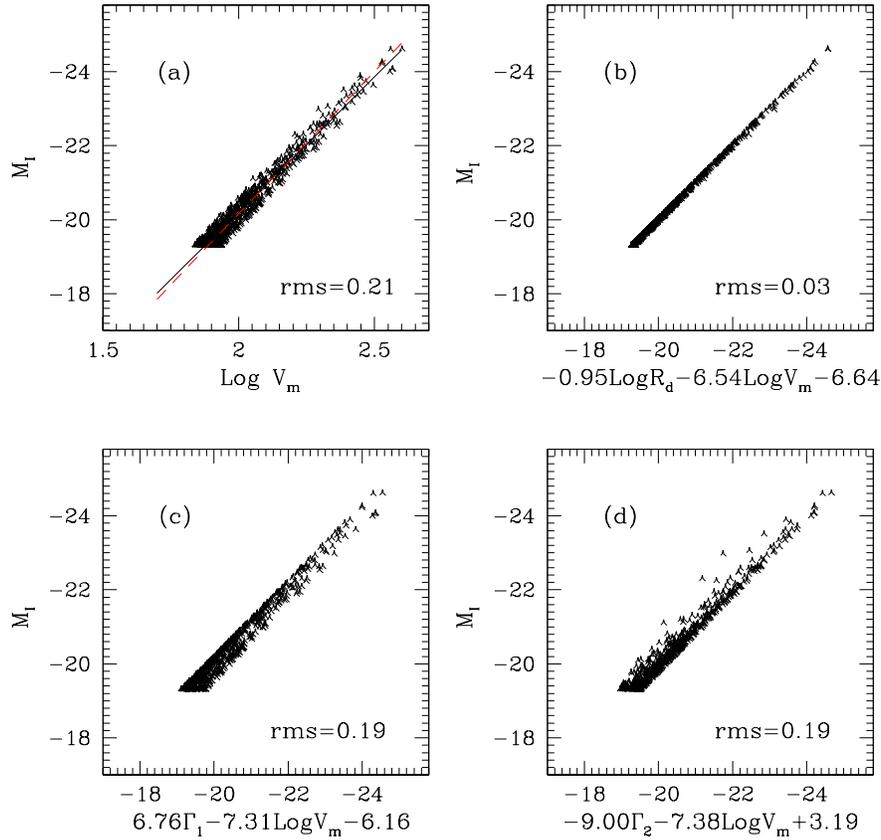}} 
\caption{The predicted TF and FP relations for Case I in Table
2.  The predicted TF relation is shown in panel
(a) together with the fit to the model prediction 
(solid line) and the fit to the observational result 
of Dale et al. (1999b) (dashed line).
 Panel (b) shows the FP with $R_d$ as the third parameter, while panel 
 (c) and (d) show the FPs based on the two different rotation-curve shape
parameters $\Gamma_1$ and $\Gamma_2$ (to be discussed in
Subsection 3.3). } 
\label{fig1}
\end{figure}

Panel (b) of Figure 2 shows the FP, while 
the corresponding TF relation is shown in panel (a)
together with
a comparison to the observational result of 
Dale et al. (1999b). We see that there is an
almost perfect FP in this case, although there is
significant TF
scatter caused by the variations of the spin parameter
$\lambda$ and
angular-momentum ratio $j_d$. The
observed slope and zero-point of the TF relation are
well reproduced, which is consistent with the result
already found in
MMW. The FP is given by $\alpha=-0.95$, $\beta=-6.54$,
with a zero-point $\gamma=-6.64$. The FP scatter in $M_I$
is only 0.03 mag,
much smaller than the predicted TF scatter ($0.21\,$mag).
Thus, the scatter in the TF relation caused by
$\lambda j_d$ can be eliminated almost entirely by the third
parameter $R_d$. So,
if the scatter in the TF relation were caused entirely by the
dispersion in the spin, the model would predict a
perfect FP for the disk population. 
However, the predicted TF scatter
($0.21\,$mag ) is lower than the observed value
of Dale et al. ($0.38\,$mag, see also Giovanelli et al. 1997), 
implying that other factors are also important in causing the
observed TF scatter.

\subsubsection{Effect of $c$}

\begin{figure}
\epsfysize=14cm
\centerline{\epsfbox{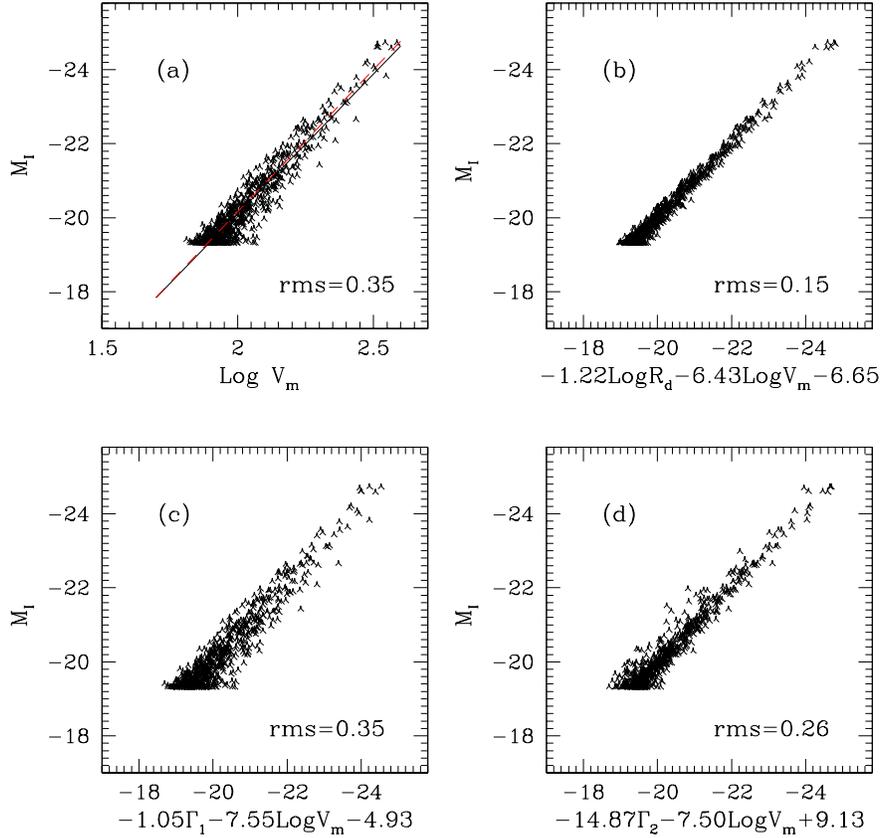}}
\caption{The predicted FP and TF relations for Case II in Table
2. The setting of the panels is the same as in Figure 2.}
\label{fig2}
\end{figure}

 Here we include the scatter caused by the
distribution of $c$, in addition to those by
$\lambda$ and $j_d$, but still keep $m_d=0.05$. The
results are shown in Figure 3. In
this case (Case II in Table 2), the observed TF
relation is also well reproduced. The predicted FP 
(with $\alpha=-1.22$, $\beta=-6.43$ and
$\gamma=-6.65$)
is close to Case I shown in Figure 1, except the
scatter is larger
($0.15\,$mag). This scatter in the FP is still
substantially smaller than that in the predicted TF
relation ($0.35\,$mag). The introduction of $R_d$ 
cannot eliminate all the scatter, because 
the FP defined by $c$ is tilted with respect to 
that defined by $\lambda$ and $j_d$ (see Table 1).

\subsubsection{Effect of $m_d$}

\begin{figure}
\epsfysize=14cm
\centerline{\epsfbox{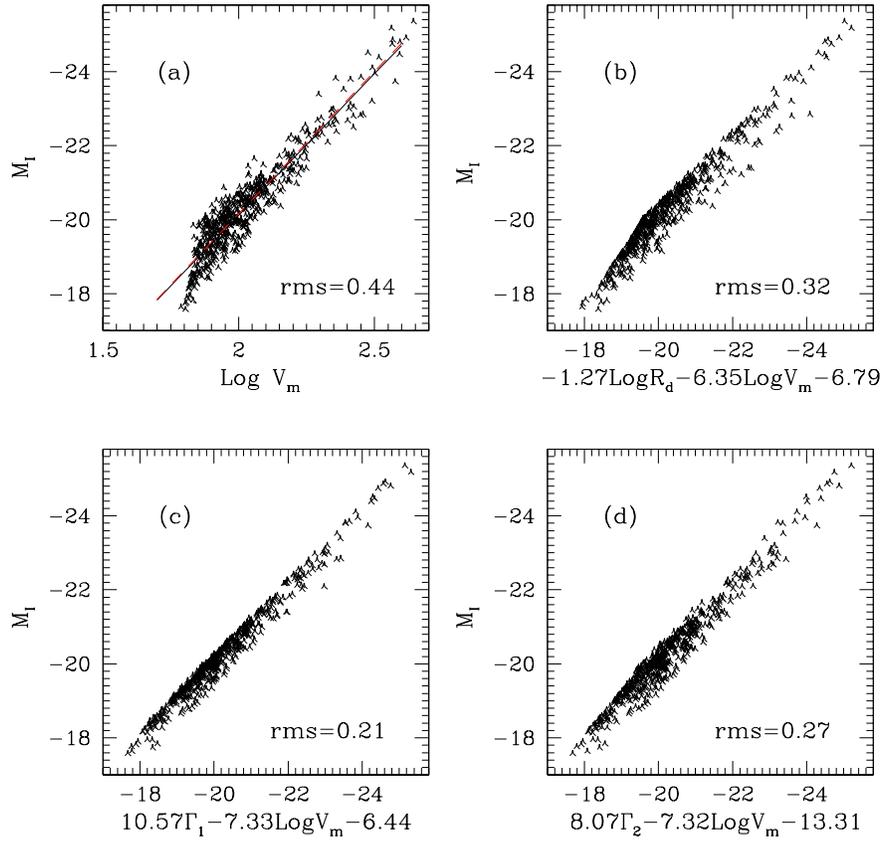}} 
\caption{The predicted FP and TF relations for Case III in Table 2.
The setting of the panels is the same as in Figure 2.}
\label{fig3}
\end{figure}

 To examine the effect of $m_d$ on the
scaling relations, we construct a Monte-Carlo sample with a fixed
concentration ($c=8.5$), with $\lambda$ and $j_d$ having
their typical distributions, and with $m_d$ randomly
drawn from 0.01 to 0.1 (Case III in Table 2). As discussed in
Subsection 2.2.4, this range 
of $m_d$ covers the values expected from any reasonable
considerations. We keep $c$ constant, in order to single out
the effect of $m_d$. The results are shown in Figure
4. In this case, the slope and zero-point of the FP
($\alpha=-1.27$,
$\beta=-6.35$ and $\gamma=-6.79$) are comparable to
those shown in Figures 2 and 3. The
scatter in the predicted TF relation ($0.44\,$mag) is much 
larger than those in  Case I and II and even larger 
than that in the observations of Giovanelli et al. (1997). 
This large predicted scatter is obviously a consequence of
the large range of $m_d$ used in this case. The scatter in
the FP relation ($0.32\,$mag) is also quite large, suggesting 
that the scatter produced by the variation of $m_d$ 
has a large tilt with respect to that caused by $\lambda$ and $j_d$
(see Table 1).
    
\subsubsection{A general case}

\begin{figure}
\epsfysize=14cm 
\centerline{\epsfbox{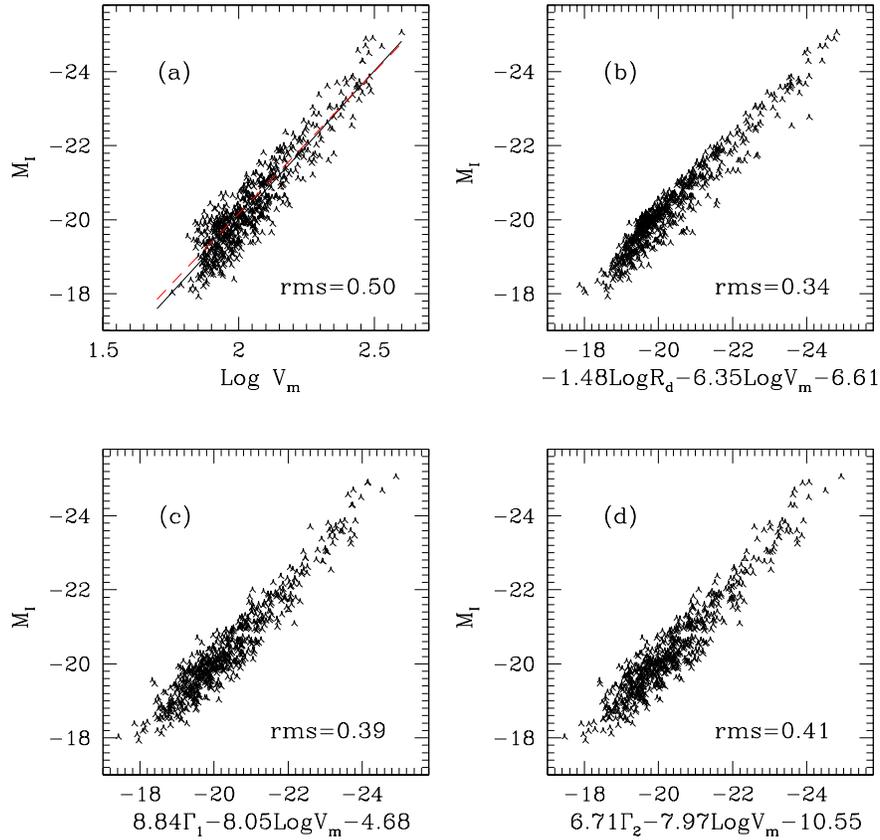}} 
\caption{The predicted FP and TF relations for Case IV in Table
2. The setting of the panels is the same as in Figure 2.}
\label{fig4}
\end{figure}

 As a summary, we consider a case where scatter
is allowed in all of the model parameters (Case IV in
Table 2). The
results are shown in Figure 5. Although generous amount of
scatter is assumed for each of the model parameters, a FP
can still be 
defined for the model galaxies, with $\alpha=-1.48$, 
$\beta=-6.35$ and $\gamma=-6.61$. 
The predicted TF relation has a scatter ($0.50\,$mag)
 larger than that observed, again mainly due to the
large range of
$m_d$ assumed. The scatter of the FP is also quite
large ($0.34\,$mag). This is not surprising, because the scatter is
dominated by the variation in $m_d$.

If we express the FP in the form $L\propto
R_d^{\alpha'}V_m^{\beta'}$,
the slopes $\alpha'$ and $\beta'$ change only little
for different cases, and the FP relation for the model
disks can be represented by
\begin{equation}\label{theplane}
L\propto R_d^{\alpha'}V_m^{\beta'}\,,
~\mbox{with}~\alpha'= 0.49\pm 0.10\,,~\beta'= 2.60\pm 0.05 \,,
\end{equation}
where the errors represent the scatter among 
different cases. Thus, the luminosity of a disk galaxy
is mainly determined 
by $V_m$, and so the TF relation is almost an edge-on
view of the FP. However, the residual dependence on $R_d$
is also significant, due to the fact that
the range of $R_d$ for the observed disks is quite large.

\subsubsection{A case with varying $\Upsilon$}

\begin{figure}
\epsfysize=14cm 
\centerline{\epsfbox{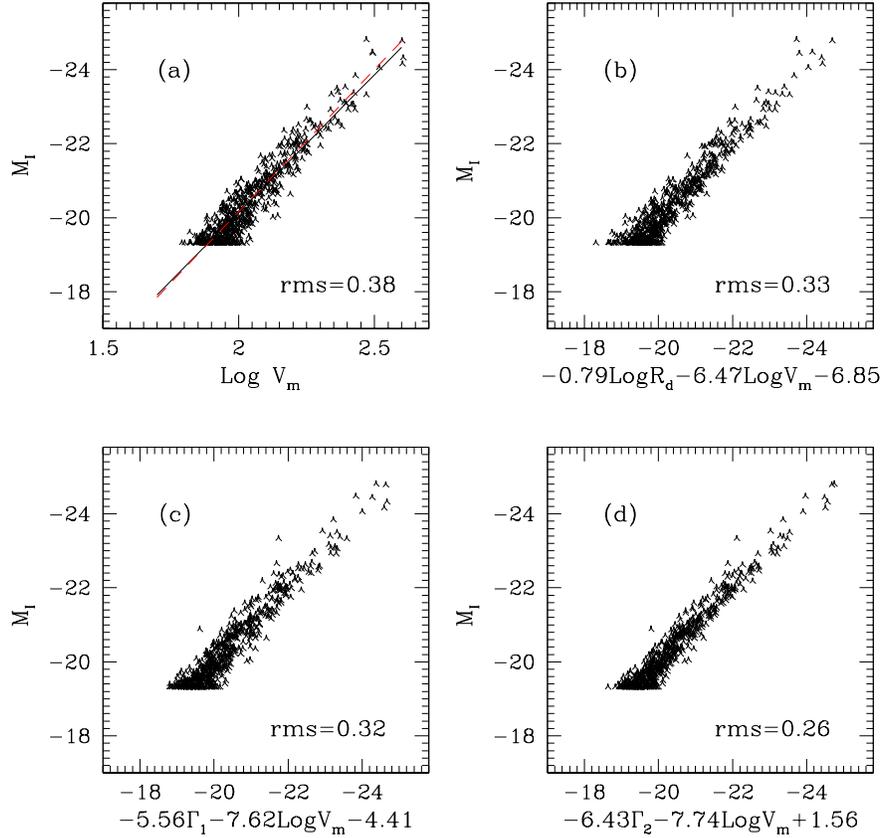}} 
\caption{The predicted FP and TF relations for Case V in Table
1. 
The setting of the panels is the same as in Figure 2.}
\label{fig5}
\end{figure}

Although we did not include the disk mass-to-light
ratio $\Upsilon$ in the list of our model parameters above,
it obviously is another parameter in the model.
In order to compare model predictions directly with 
observations, we need to convert the predicted disk
mass to a disk luminosity,
which requires an assumption about the disk mass-to-light
ratio. Scatter in the disk mass-to-light ratio therefore
also gives rise to scatter in both the TF and the FP
relation. Clearly, the induced scatter is along the $M_I$ axis,
with a $rms$ (in terms of the luminosity) exactly the
same as
that in the mass-to-light ratio. 
   
The exact value of $\Upsilon$ for a galaxy depends
on its star formation history and is in general 
difficult to model accurately. As an example,
we consider a case (Case V in Table 2)
where the disk mass-to-light ratio
is assumed to be 
\begin{equation}
\Upsilon=1.7h(m_d/0.05)
\end {equation}
and all other parameters are assumed to be the same as
in Case IV. In this case, the effect of $m_d$ on the magnitude
$M_I$ is eliminated, and so the TF scatter caused by $m_d$
is only through its effect on the rotation curve.
The results are shown in Figure 6. The predicted TF
relation has scatter smaller than that in Case IV because the
effect of $m_d$ on $M_I$ is eliminated [see Equation (6)]. 
The scatter of the predicted FP ($0.33\,$mag) is close to
that of the predicted TF relation ($0.38\,$mag), implying that
the scatter in the TF relation caused by $m_d$ through
disk action cannot be effectively reduced by
introducing $R_d$ as the third parameter. 

  From the results presented above we can conclude
that variations in the mass ratio $m_d$ and in the
mass-to-light ratio may be important 
sources of scatter in the predicted TF
and FP relations. If the variations in $m_d$ and
$\Upsilon$
are small, the scatter in the predicted FP relation
may be significantly lower than that in the TF relation. On
the other hand, if the scatter in the observed TF relation is
mainly due to the variations in $m_d$ and $\Upsilon$, the
introduction of the
third parameter $R_d$ will not reduce the scatter
significantly.
Thus, by comparing the scatter in the TF relation with
the scatter in the FP relation, one may hope to find the 
main sources of scatter in the two scaling relations. We will come
back to this issue in Section 4.

\subsection{Results Based on Rotation Velocities at
Other Radii}

 Although the maximum rotation velocity ($V_m$) is
commonly used in defining the TF relation
(mainly for observational reasons, because the
inclination-corrected 21 cm line width of a disk
galaxy
is assumed to be twice its maximum rotation velocity),
there is no {\it a priori} reason against using
rotation velocities at other radii. Such definition is
possible
for disk galaxies with measured rotation curves.
Since the disk contribution to the rotation curve
may change with radius (for example, the rotation
curve of a galaxy at large radius is expected to
be dominated by dark matter while the disk
contribution
may be significant in the inner region), analyses
of the TF relation (or the FP relation) using rotation
velocities at different radii may provide more
information
on the mass components of disk galaxies.

\begin{figure}
\epsfysize=14cm \centerline{\epsfbox{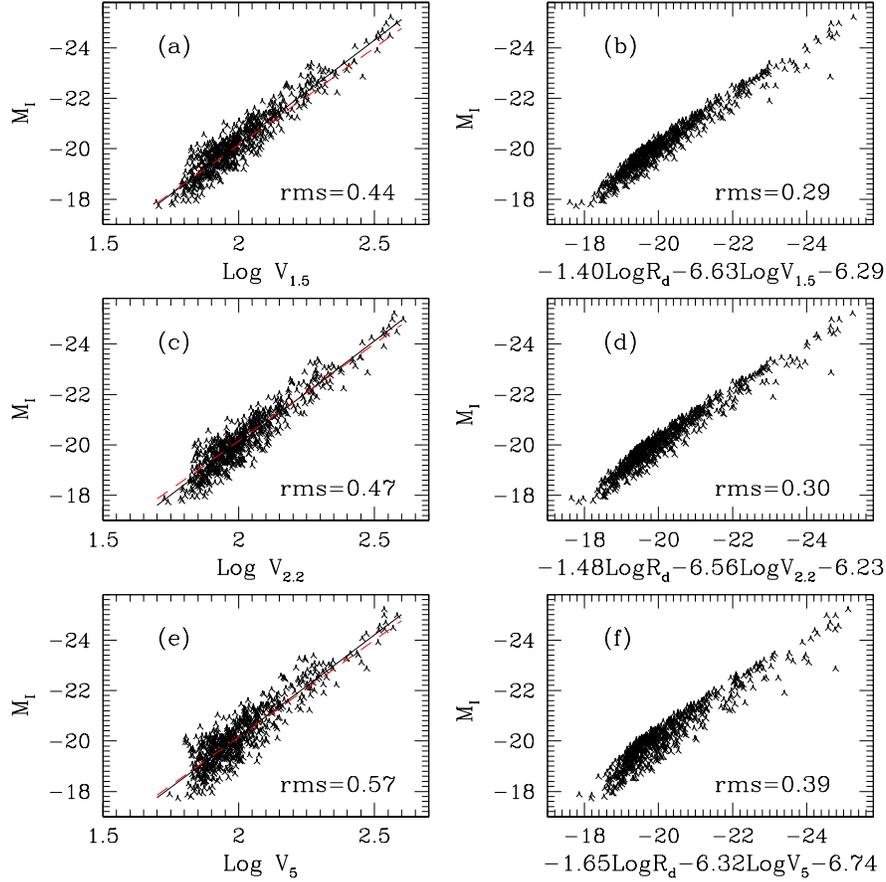}} 
\caption
{The predicted TF
and FP relations based on different rotation
velocities. The right panels show the  
FPs with $R_d$ as the third parameter, using 
respectively $V_{1.5}$, $V_{2.2}$ and
$V_{5}$ (all in units of $\kms$)
as the characteristic rotation velocity. 
The left panels show the corresponding TF relations. 
The solid lines in the TF panels are the fits to the
model predictions, while the dashed lines represent the
relation observed by Dale et al. (1999b).}
\label{fig6}
\end{figure}

 Observationally, there are some attempts to define
the TF relation using rotation velocities other than
the maximum rotation velocity, in the hope of getting smaller
scatter. For example, Courteau (1997) found an optimal 
TF relation using
rotation velocities at $2.2 R_d$, while there are also
claims that the TF relation is optimized if 
rotation velocities at smaller radii are used
(see Willick 1999 and references therein).
Interestingly, almost all of these analyses 
showed that the scatter in the TF relation becomes 
larger when the amplitudes of the rotation curves
at large radii are used.

In this section, we use our theoretical model to
examine the
effect of using different rotation velocities on the
TF and FP relations. Specifically, we choose three
representative radii,
$R=5R_d, 2.2R_d, 1.5R_d$,  to define the
characteristic rotation
velocities ($V_{5}$, $V_{2.2}$ and $V_{1.5}$). For
most of our model
rotation curves, the maximum rotation velocity is
reached at
$R\sim 3R_d$ (MMW), and so $V_{5}$ represents the
rotation velocity
well beyond the peak of a rotation curve. $V_{2.2}$ is
chosen because the rotation velocity of a pure exponential
disk peaks at
$R\approx 2.16R_d$, while $V_{1.5}$ is chosen to
represent the inner
part of the rotation curve.

 We construct a Monte-Carlo sample in the same way as
 described in Subsection 2.3. The distributions of all
the model parameters 
are assumed to be the same as the general case (Case IV) in Table 2.
The results are shown in Figure 7 and summarized
in Table 3.

\begin{table}
\caption{The same as Table 2 except that other   
rotation velocities are used instead of $V_m$.}
\begin{tabular}{lcccccccccc} \hline
Case & $\lambda$ & $j_d$ & $c$ & $m_d$ &  $\Upsilon$ &
$\alpha$ & $\beta$
& $\gamma$ 
& $\sigma_{\rm FP}$  &  $\sigma_{\rm TF}$   \\
\hline
$V_{1.5}$ & eq.\,(7) & eq.\,(10) & eq.\,(9) & 0.01-0.1 
&$1.7h$& $-1.40$ & $-6.63$ & $-6.29$ & 0.29 & 0.44 \\ 
$V_{2.2}$ & eq.\,(7) & eq.\,(10) & eq.\,(9) & 0.01-0.1 
&$1.7h$& $-1.48$ & $-6.56$ & $-6.23$ & 0.30 & 0.47 \\ 
$V_{5} $ & eq.\,(7) & eq.\,(10) &   eq.\,(9) & 0.01-0.1
&$1.7h$& $-1.65$ & $-6.32$ & $-6.74$ & 0.39 & 0.57 \\  
\hline
\end{tabular}
\end{table}

As shown in Figure 7, the model predictions are
consistent with the observations that the use of rotation 
velocities at larger radii generally leads to  larger 
scatters in the TF relations. This result can be understood 
as follows. At very large radii, the disk contribution 
to the rotation curve is negligible, and so the 
variation in $m_d$ does not affect the
rotation velocities very much. Since the 
luminosity of a disk is directly proportional to $m_d$
(for a constant mass-to-light ratio), the variation in $m_d$
affects the TF scatter directly. This effect is reduced 
in the cases where rotation velocities in the inner part 
(e.g. $V_{2.2}$ and $V_{1.5}$) are used, because of
the increasing halo-disk interaction (Mo \& Mao 2000; 
Navarro \& Steinmetz 2000). 
We have also made calculations for models where 
$m_d$ is kept constant. In such cases, the TF 
scatter is almost independent of the definition of the 
characteristic rotation velocity. Thus, the scatter in 
the TF relations defined at different rotation velocities 
may be used to constrain the variation of the mass ratio
$m_d$.
 
Similar to the TF relations, the scatter in the FP relation
is also larger when the rotation velocities at larger radii are used.
This is because $R_d$ cannot effectively reduce the TF scatter 
caused by $m_d$.

\subsection{Rotation-Curve Shape as the Third Parameter}

Since rotation curves of individual galaxies have
different shapes, it is possible that the TF scatter 
is correlated with the rotation-curve shape. 
Persic \& Salucci (1990) considered this possibility
from observational data. In the theoretical model
we are considering here, the shapes of rotation curves
can be affected by all the four model parameters, and
so the TF scatter must be correlated to the
rotation-curve shape to some degree. In this
subsection, we
carry out a quantitative analysis of such correlation.
To do this, we use two parameters similar to those 
defined by Persic, Salucci \& Stel (1996),
\begin{equation}
\Gamma_1=\left(\frac{d\,\Log\,V}{d\,\Log\,R}\right)_{2.2R_d},
~~~      
\Gamma_2=\frac{V_{5}}{V_{2.2}},
\end{equation}
to represent the rotation-curve shapes in the 
inner and outer regions. We investigate the FP in the space 
spanned by $M_I$, $\Log(V_m)$ and $\Gamma_1$ ($\Gamma_2$)   
for all the cases considered above.
The simulated FP is fitted to the relation
\begin{equation}
M_I=\alpha\Gamma + \beta\Log V_m (\kms)+\gamma,
\end{equation}
where $\Gamma$ is either $\Gamma_1$ or $\Gamma_2$.
The predicted FPs are shown 
in panel (c) and panel (d) in each of Figures 2--6,
and the fitting results are summarized in 
Table 4.

\begin{table}
\caption{The predicted FP relations where the rotation-curve
shape parameters $\Gamma_1$ and $\Gamma_2$ are adopted as the 
third parameter. Column 1 lists the name of
each case  as listed in Table 2. Columns 2 to 5 give 
the fit results for the FP 
in the $M_I$-$\Log V_m$-$\Gamma_1$ space,
while columns 6 to 9 are those for the FP in
the $M_I$-$\Log V_m$-$\Gamma_2$ space.}

\begin{tabular}{lrrrrrrrrr} \hline
& &$\Gamma_1$& & & & &$\Gamma_2$\\
\cline{2-5}
\cline{7-10}
Case 
&$\alpha$ & $\beta$ & $\gamma$&$\sigma_{\rm FP}$  &
&$\alpha$ & $\beta$ & $\gamma$&$\sigma_{\rm FP}$
\\
\hline
I 
& 6.76   & $-7.31$ & $-6.16$ &0.19& &$-9.00$ & $-7.38$ & 3.19 & 0.19 \\
II 
& $-1.05$ & $-7.55$ & $-4.93$ &0.35& &$-14.87$& $-7.05$ & 9.13 & 0.26  \\
III
& 10.57   & $-7.33$ & $-6.44$ &0.21& & 8.07  & $-7.32$ &$-13.31$&0.27\\
IV 
& 8.84 & $-8.05$ & $-4.68$ & 0.39 & &6.71& $-7.97$&$-10.55$& 0.41 \\
V & $-5.56$ & $-7.62$ & $-4.41$ & 0.32 & &$-6.43$&$-7.74$&1.56& 0.26 \\
VI & $-0.28$ & $-9.97$ & $0.44$ & 0.35 & & $-8.57$& $-10.69$&1.94& 0.34 \\
\hline

\end{tabular}
\end{table}

These results show that the introduction of the shape 
parameter of rotation curves as the third parameter can 
reduce the scatter of the TF relation, but the two  
parameters considered here are less effective than 
$R_d$ in all the cases except in Case III where 
the TF scatter is effectively reduced by the 
introduction of $\Gamma_1$ or $\Gamma_2$. This suggests that 
these shape parameters are effective in reducing 
the scatter caused by $m_d$ (due to disk action)
but not so much in reducing the scatter 
caused by $\lambda$$j_d$ and $c$. 

  Unfortunately, $\Gamma_1$ and $\Gamma_2$ are both more 
difficult to measure 
than $R_d$ from observations, and it is not yet
possible to obtain such a FP from observational data. 
Notice that the slope $\alpha$ associated with
the two shape parameters is very sensitive to the change 
of model parameters, because the ranges of these parameters 
are relatively small. 

\begin{figure}
\epsfysize=14cm \centerline{\epsfbox{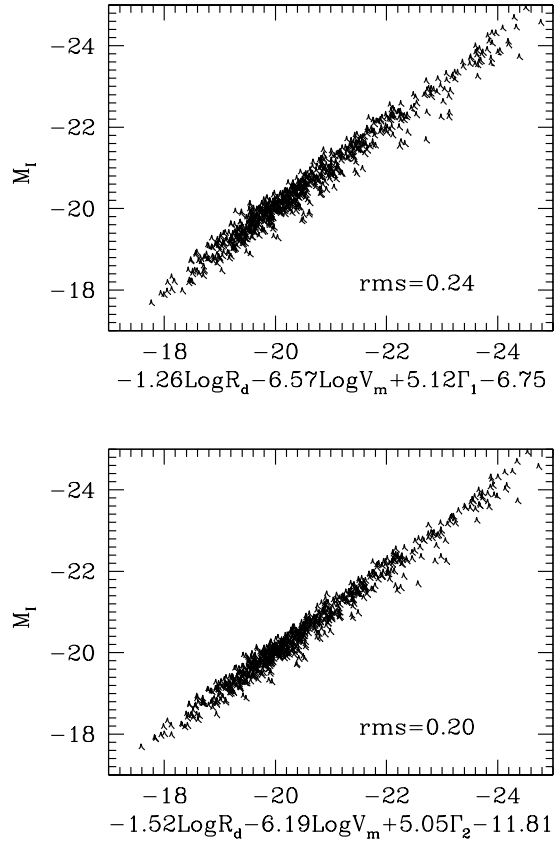}}
\caption
{The edge-on view of the FP in the space spanned by 
$M_I$, $\Log R_d$, $\Log V_m$, $\Gamma$ (or $\Gamma_2$)
for Case IV. Notice that the scatter in $M_I$ 
is reduced significantly relative to that 
shown in Figure 5.} 
\end{figure}

 From the above discussion we see that the disk scale-length
$R_d$ is effective in reducing the TF scatter 
caused by $\lambda j_d$ (Figure 2) and $c$ (Figure 3),
while the rotation-curve shape parameters are more effective 
in reducing the scatter caused by $m_d$ (Figure 4). It is therefore 
interesting to see what happens if both $R_d$ and one of the 
rotation-curve shape parameters are introduced  
to define a plane in the four-dimensional space
spanned by $M_I$, $\Log R_d$, $\Log V_m$, and $\Gamma_1$ 
(or $\Gamma_2$). The results are shown in Figure 8 for Case IV. 
As expected, the scatter is significantly reduced.

\subsection{A case where $m_d$ changes with $V_c$}

In reality, the value of $m_d$ may depend
systematically on halo circular velocity, as is the case 
if feedback effects can eject gas more effectively from smaller
haloes (e.g. White \& Frenk 1991). To model this effect,
we construct a sample where $m_d$ is assumed to change 
with $V_c$ as 
\begin{equation}
m_d=0.1\left[1+\left(\frac{150\kms}
{V_c}\right)^2\right]^{-1}(1+\delta)\,,
\end{equation}
where $\delta$ is a random number between 
$-0.1$ and 0.1. Other parameters are chosen the same 
as in Case IV. This case is listed as Case VI in Tables 2
and 4, and the results are shown in Figure 9.
\begin{figure}
\epsfysize=14cm \centerline{\epsfbox{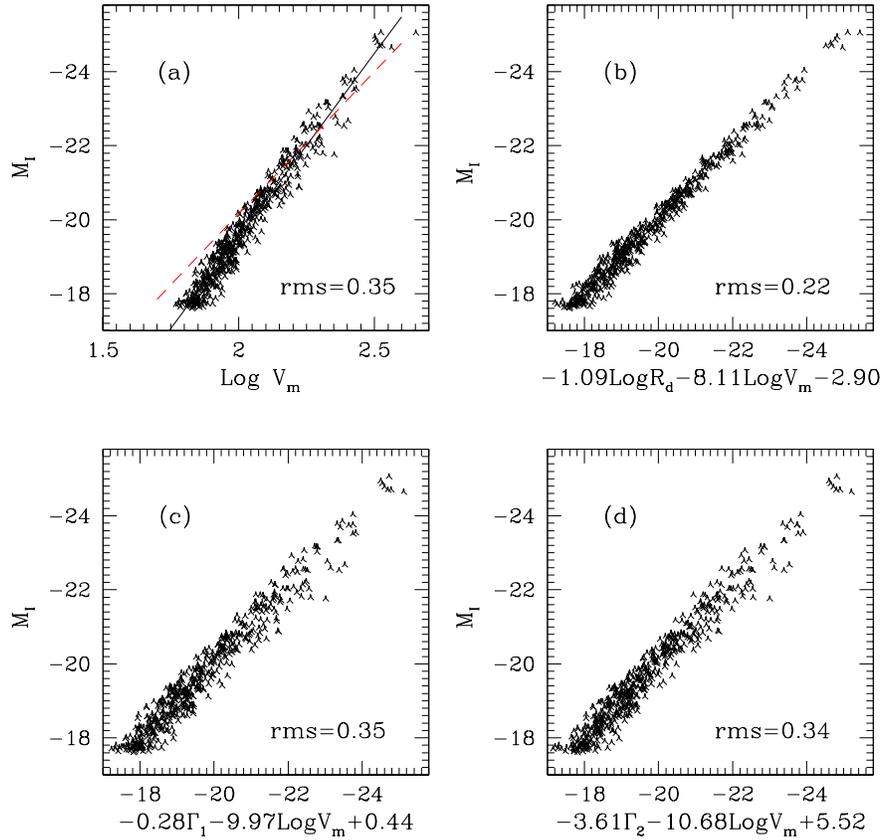}}
\caption
{The predicted FP and TF relations for Case VI in Table 2. 
The setting of the panels is the same as Figure 2}
\end{figure}
In this case the TF relation is  
$M_I=-9.75\Log V_m-0.03$ [the solid line 
in panel (a) of Figure 9], which has a much steeper 
slope ($\beta\approx 3.8$)
than all other cases we have considered. 
Correspondingly, the slope $\beta$ in the
FP relation (with $R_d$ as the third parameter) 
is also steeper, but the value of $\alpha$ does
not change much (see Table 2). The introduction of $\Gamma_1$ (or $\Gamma_2$) 
as the third parameter does not help much in reducing 
the TF scatter because, with the assumed small scatter 
in $m_d$, the TF scatter is dominated by the variations
in $\lambda$$j_d$ and $c$.

\begin{figure}
\epsfysize=14cm \centerline{\epsfbox{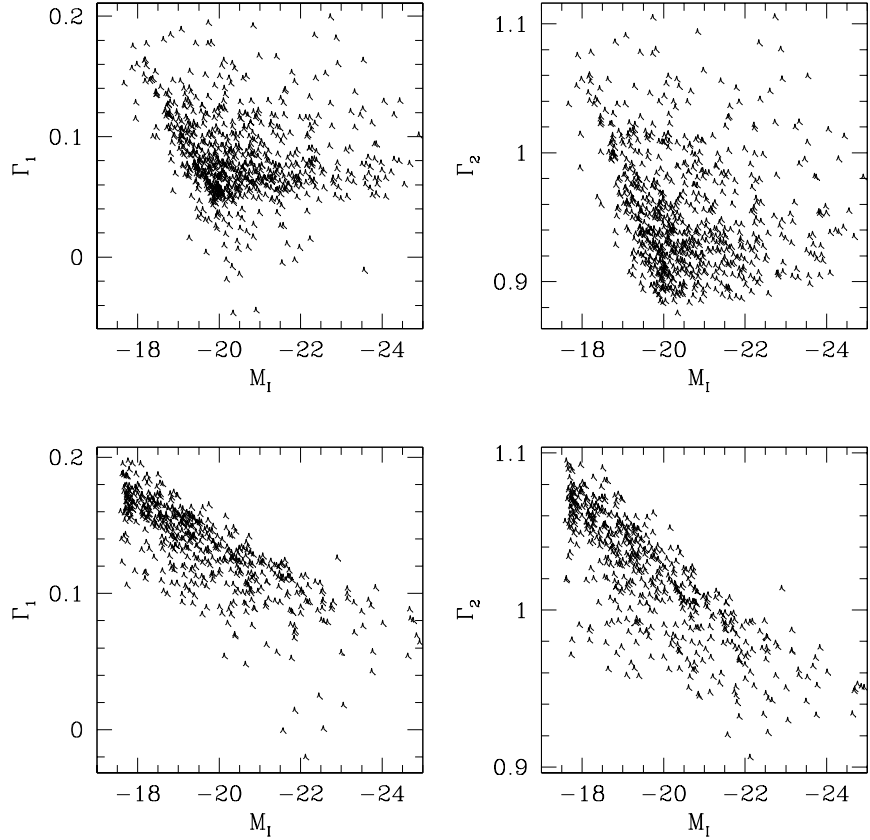}}
\caption
{The correlation between $M_I$ and the rotation-curve shape 
parameters, $\Gamma_1$ and $\Gamma_2$. 
The top two panels are for Case IV while the
lower two panes are for Case VI where $m_d$ is correlated with $V_c$
as in Equation (18).}
\end{figure}

 Since the rotation-curve shape parameters are correlated 
with $m_d$, it is interesting to see if these parameters 
can be used to reveal the trend of $m_d$ with $V_c$.
To do this, we examine the correlation between $M_I$ 
and $\Gamma_1$ (or $\Gamma_2$) for Case VI and compare the results
with those for Case IV. The results are shown in 
Figure 10. As one can see, if 
$m_d$ is correlated with $V_c$, then there is a strong correlation 
between $M_I$ and $\Gamma_1$ (or $\Gamma_2$), in the
sense that more luminous galaxies have smaller
$\Gamma_1$ (or $\Gamma_2$). Such correlations 
are in fact observed by Persic, Salucci \& Stel (1996),
but the data are still too uncertain to give any meaningful 
constraints on the model. 

\section{Comparison with Preliminary Observational
Results} 
\label{observations}

\begin{figure}
\epsfysize=14cm \centerline{\epsfbox{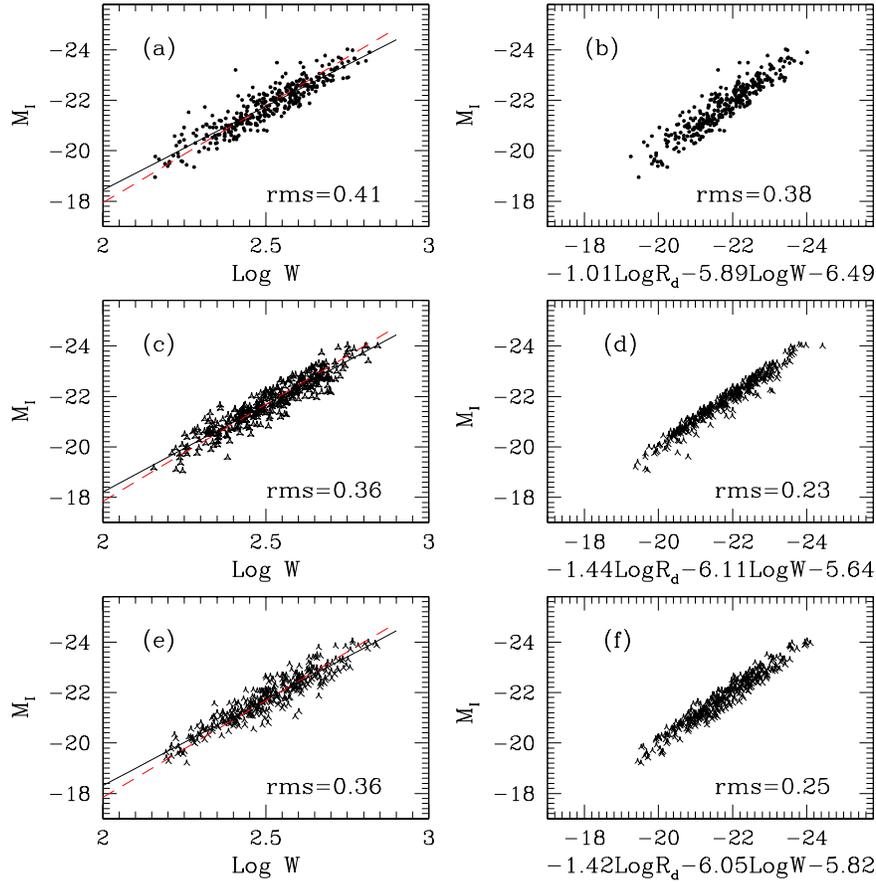}}
\caption
{The observed and simulated FP and TF relations based on the data 
from Dale et al.(1999a). Panel (a) shows the  FP while panel (b) shows the TF relation.
Panels (c), (d), (e) and (f) show  the FPs and TF relations for 
two simulated samples which have the same distributions in $L$
and $R_d$ as the observed sample. Panels (c) and (d) show the 
theoretical predictions of Case IV. Panels (e) and (f) show the
theoretical predictions of a sample with the same parameter
as Case II except that the mass-to-light ratio
$\Upsilon$ is assumed to be random between $1h$ and $2h$. 
 The solid lines in three TF panels are the direct fit to the data point, 
 while the dashed lines show the result obtained
by Dale et al. (1999b) after bias correction. 
} 
\label{observfig}
\end{figure}

 Having seen the theoretical expectations for the
scaling relations of spiral galaxies, one is obviously
interested in seeing what observations actually show.
Here we present some preliminary results without going into
the details of the observational selection effects,
leaving a more sophisticated analysis to a future paper.

The data we use are from Dale et al. (1999a), which
presents TF observations for 35 rich Abell clusters of galaxies
in the $I$-band, along with the exponential scale-lengths $R_d$.  
We simply use the published data without correcting for
any bias. 
 Panel (a) and (b) of Figure \ref{observfig}  show the results. 
The FP for the observed galaxies can be described as
\begin{equation}
    M_I=-1.01\Log R_d(\kpc)-5.89\Log W(\kms)-6.49,
\end{equation}
where $W$ is the velocity width, which we assume to be twice
$V_m$. 
The index on $R_d$ is comparable to that given by the
models presented above, but the index on $W$ (or $V_m$) 
is lower than the model predictions. This lower value is 
almost certainly due to observational bias. 
In order to show this, we construct a sample 
with the same model parameters as in the general case (Case IV) and
select model galaxies with the same luminosity and size
distributions as the observed sample so as to reproduce the selection effect. 
The results are shown in panels (c) and (d) of
Figure 11. As one can see, the TF and FP slopes for
this sample are quite close to the observed values. 
Note that the scatter is smaller than that of Case IV in Figure 5, 
indicating the selection effect is important. 
As another example, we show in
panels (e) and (f) the results for a sample with the same model 
parameters as in Case II except that $\Upsilon$ has a random 
distribution between $1h$ and $2h$ and with galaxies 
also selected according to the observed luminosity and size 
distributions. We see that the scatter allowed in 
$\Upsilon$ (a factor of two) is much smaller than that
in $m_d$ (a factor of ten), again because disk action
reduces TF scatter (Mo \& Mao 2000). In both cases, the
scatter in the FP is significantly smaller than that
observed. Unfortunately, it is unclear how seriously 
these discrepancies should be taken. The bias correction 
required may be more complicated than what is assumed here. 
In particular, the observed sample is for cluster galaxies, 
and so systematic bias may also arise from some 
environmental effects which are not modelled in the theory.

\section{Discussion and Summary}

In this paper, we use current theory of disk galaxy
formation to study whether a FP-like relation is expected for
spiral galaxies.
After examining in detail how the changes in model
parameters affect
the scaling relations of disk galaxies, we find that
the fundamental properties of disk galaxies are
generally concentrated into a plane, with the TF relation
representing an almost edge-on view. 
We made a systematical search for the third parameter 
which may correlate with the TF scatter.  
We find that the disk scale-length $R_d$ 
(or surface brightness) as the third parameter  
can effectively reduce the scatter of TF relation, especially that caused by 
the variations of halo spin and concentration.  
The rotation-curve shape as the third parameter can be used
to reduce the scatter caused by $m_d$.  
For the various cases we analyzed, 
the FPs in the $(\Log L,\Log V_m, \Log R_d)$-space 
are quite similar and well represented by equation
(\ref{theplane}). This relation is consistent with the preliminary
results we obtain from observational data. 

 We also discuss  the possibility of using other
characteristic velocities [e.g. $V_{5}$, $V_{2.2}$
and $V_{1.5}$] instead of $V_m$ as the velocity parameter 
in the TF and FP relations.
The theory is consistent with the observations that
the scatter in the TF relation is generally larger when rotation
velocities at larger radii are used. 

There are, however, a number of uncertainties in
the theoretical model, which must be taken into
account when comparing models with observations.
Real galaxies may be much more complicated than our
simple model implies: galaxy disks may not be
perfectly exponential, 
the mass distribution in dark haloes may be
non-spherical and clumpy and so the rotation curves may not be
smooth, and the existence of galactic bulges may affect both 
the rotation curve and the luminosity profile.  
Furthermore, the assumption of a constant disk
mass-to-light ratio (in the $I$-band), although consistent with current
observations, is clearly unrealistic, because the
mass-to-light ratio of a galaxy depends on its star 
formation history. None of these issues are easy to 
resolve, but we hope the present paper can provide 
some theoretical guidelines for the search of scaling
relations for disk galaxies.

\section*{Acknowledgments}
The authors thank Shude Mao, Frank van den Bosch and Simon D. M. White 
for carefully reading the manuscript and useful suggestions. 
SS and CS acknowledge the financial support of
MPG for visits to MPA. SS thanks R. Chan and D. Chen for
helpful discussion. This project is partly supported by the
Chinese National Natural Foundation, the WKC foundation and
the NKBRSF G1999075406. 

{}
\end{document}